\documentclass{JHEP3}
\usepackage{epsfig}
\usepackage{graphicx}
\usepackage{dcolumn}
\usepackage{bm}
\newcommand{\be}{\begin{equation}}
\newcommand{\ee}{\end{equation}}
\def\bea{\begin{eqnarray}}
\def\eea{\end{eqnarray}}

 \def\be{\begin{equation}}
\def\ee{\end{equation}}
\def\bea{\begin{eqnarray}}
\def\eea{\end{eqnarray}}

\def\lesssim{\mathrel{\hbox{\rlap{\hbox{\lower4pt\hbox{$\sim$}}}\hbox{$<$}}}}
\def\gtrsim{\mathrel{\hbox{\rlap{\hbox{\lower4pt\hbox{$\sim$}}}\hbox{$>$}}}}

\title{Flux vacua as 
supersymmetric attractors}

\author{ Renata Kallosh
\\
    Department of Physics, Stanford University, Stanford, CA 94305,
USA and \\
Kyoto University, Yukawa Institute, Kyoto, 606-8502, Japan}
\received{\today}       
 \preprint{YITP-05-48 \\  ~hep-th/0509112\\ September 15, 2005}

\abstract{We derive algebraic attractor equations describing supersymmetric flux vacua of  type IIB string theory in terms of a doublet of the  3-form fluxes, $F$ and $H$. These equations are similar to the attractor equations for moduli fixed by the charges near the horizon of the supersymmetric black holes. 
}

\begin{document}

\section{Introduction}

The supersymmetric attractor equations \cite{Ferrara:1995ih}-\cite{Ferrara:1996um} have been discovered in the context of the theory of BPS black holes. It was found there that the scalar fields near the black hole horizon have a fixed point behavior defined by the black hole electric and magnetic changes.    The purpose of this note is to show that there is a direct relation between the mathematical structure established for supersymmetric black hole attractors and a particular class of flux vacua. 

A significant progress in investigation of this question was achieved by Moore in \cite{Moore:2004fg}.\footnote{For an early attempt to relate the minimum of the superpotential to the $SL(2,Z)$ black hole attractors see \cite{Kallosh:1996xa}; for a more recent discussion on flux vacua and attractors see \cite{Curio:2000sc},\cite{Ooguri:2005vr}.} However, his investigation revealed a problem which did not allow an immediate generalization of the results for black hole attractors to the study of flux vacua. He  pointed out that in type IIB string theory compactified on a CY$_3$ manifold the flux vacua are described by the complex non-integral $\tau$-moduli-dependent 3-form flux $G_3= F_3-\tau H_3$. The appearance of the field $\tau$ in $G_{3}$  is the source of a problem when one tries to use the black hole attractor equations based on a set of real magnetic and electric charges.
It was  suggested in \cite{Moore:2004fg} that this issue is easier to resolve in the context of M-theory compactified on a CY$_4=$CY$_3\times T^2$ manifold. Moore has developed these ideas for a particular example when CY$_3= K3\times T^2$. 

In this paper we will find the attractor equations for the general case of type IIB string theory compactified on CY orientifolds. We choose O3/O7 orientifolds to be specific, as in the GPK-KKLT vacua \cite{Giddings:2001yu}, \cite{Kachru:2003aw}.

We will show that for generic CY$_3$ manifold with O3/O7 planes one can rewrite the standard {\it differential} equations for flux vacua in IIB string theory, $DW=0$ with  $W\neq 0$, as  {\it algebraic} attractor equations relating fluxes to fixed moduli,  including all complex structure moduli as well as the axion-dilaton:
\begin{eqnarray}
DW=0 , \quad  W\neq 0 \hskip 1 cm \Rightarrow \hskip 1 cm p^\Lambda  = i \bar Z L^\Lambda - i  Z \bar L^\Lambda , \quad q_\Lambda  = i \bar Z M_\Lambda - i  Z \bar M_\Lambda \ .
\label{attr}
\end{eqnarray}
Here $p, q$ are quantized 3-form fluxes and the relevant covariant holomorphic central charge is $Z= e^{K/2} W$.  The covariantly holomorphic section $ (L^\Lambda, M_\Lambda) $ will be presented below. For the black hole attractor equations (\ref{attr}) to be valid in flux vacua one has to require in addition  that the second derivative of the central charge is vanishing, $D_{i}D_{j} Z=0$, see \cite{Kallosh:2005ax}. For the BPS  black hole case in N=2 supergravity this condition is always satisfied, see for example eq. (55) in \cite{Ferrara:1997tw}.

An unusual situation here is that we are looking for N=1 flux vacua  with the N=1 effective potential 
\begin{equation}
  V= e^K(|D W|^2-3 |W|^2) \ .
\label{N1pot}
\end{equation}
The N=1 potential (\ref{N1pot}) is based on the GVW superpotential \cite{Gukov:1999ya}
\begin{equation}
W=\int G_3\wedge \Omega
\label{GVW}
\end{equation}
The vacua of our interest are defined by equations $DW=0$ and $W\neq 0$ so that the potential at the critical point with $\partial V=0$ is given by
$
  V_{cr}= -3 e^K|W|_{cr}^2
$.
If the effective metric in 4d space time is of AdS type, this is a supersymmetric vacuum; if the metric is Minkowski, this is a non-supersymmetric one. Either way, our problem here is to show that instead of solving {differential} equations on the superpotential, $DW=0$, the corresponding critical point of the potential can be found   by solving {algebraic} attractor equations relating fluxes to fixed moduli. Note that this is by no means a property of a general N=1 supergravity. An important ingredient of our construction  follows from compactified type IIB string theory on CY orientifold: on CY manifold we  get N=2 supergravity upon compactification and  orientifolding, which truncates it to N=1 but does not change the N=2 properties of the geometry of the moduli space surviving orientifolding. After orientifolding, some fields are truncated from the theory. However, for the remaining moduli  one still finds the formal structure of N=2 theory, as in special geometry \cite{deWit:1984pk}.

In effective N=1 supergravity obtained by compactification of string theory on CY orientifold one finds that there exists a covariantly holomorphic section from which the K\"ahler potential $K$ is constructed.  The covariantly holomorphic central charge  $Z=e^{K/2}W$ is a symplectic invariant, which transforms under K\"ahler transformations as a phase. 
Therefore $|Z|^2$ and $|DZ|^2$ are both symplectic and K\"ahler invariant. Thus the  geometry of the moduli space, as well as the symplectic invariance of the central charge, remains in the effective N=1 supergravity, which explains the presence of  attractor equations in flux vacua.
The potential in (\ref{N1pot}) can be therefore presented in the form
\begin{equation}
  V= |DZ|^2- 3|Z|^2 \ ,
\label{pot}
\end{equation}
where $ Z(z, \bar z, q,p) \equiv  (L^\Lambda
q_\Lambda -
M_\Lambda p^\Lambda)$. Here $p,q$ are integer 3-form fluxes and  $ (L^\Lambda, M_\Lambda) $ is a covariantly holomorphic section for IIB string theory compactified on CY orientifold.

\section{ Attractors and Special Geometry} 

Here we give a short overview of attractors and special geometry following \cite{Ferrara:1995ih}-\cite{Ferrara:1996um}, \cite{deWit:1984pk}, where we will comment on particular features related to flux vacua. 
A special K\"ahler manifold can be defined  by constructing flat
symplectic
bundle of dimension $ 2n+2$ over K\"ahler-Hodge manifold with
symplectic
section defined as
$$
V=(L^\Lambda, M_\Lambda) ,\qquad \Lambda = 0,1,...n\ ,
$$
where $(L,M)$ obey the symplectic constraint
$
i(\bar L^\Lambda M_\Lambda - L^\Lambda \bar M_\Lambda)=1
$ and
 $L^\Lambda(z, \bar z) $   and $M_\Lambda(z, \bar z)$ depend on
scalar fields
$z,\bar z$, which are the coordinates of the ``moduli space."
$ L^\Lambda$ and $M_\Lambda$ are {\it covariantly
holomorphic} (with respect to the K\"ahler connection), e.g.
$$
D_{\bar k} L^\Lambda = (\partial_{\bar k} - {1\over 2} K_{\bar
k})L^\Lambda =0
\ ,
$$
 where K is the K\"ahler potential. Symplectic invariant  form of
the K\"ahler
potential can be found from this equation by introducing the {\it holomorphic
section}  $(X^\Lambda (z), F_{\Lambda}(z))$:
$$
L^\Lambda = e^{K/2} X^\Lambda \ , \qquad M_\Lambda = e^{K/2} F_\Lambda\ ,
\qquad
(\partial_{\bar k}
X^\Lambda  = \partial_{\bar k} F_\Lambda=0) \ .
$$
The K\"ahler potential is
$
K = -\ln i( \bar X^\Lambda F_\Lambda -   X^{\Lambda } \bar F_\Lambda).
$
The K\"ahler  metric is given by $g_{k\bar k} = \partial_k
\partial_ {\bar k}
K$. Finally, from special geometry one finds that there exists a complex
symmetric $(n+1)\times (n+1) $ matrix ${\cal N}_{ \Lambda
\Sigma}$ such  that
\begin{equation}
M_\Lambda = {\cal N}_{ \Lambda \Sigma} L^\Sigma \ , \qquad
{ \rm {Im\,}}{\cal N}_{ \Lambda \Sigma} L^\Lambda  \bar L^\Sigma
=-{1\over 2} \ ,
\qquad
D_{\bar i} \bar  M_\Lambda = {\cal N} _{\Lambda \Sigma} D_{\bar i}  \bar
L^\Sigma \ .
\label{Ncal}
\end{equation}
In the case of N=2 supergravity this matrix is a metric in the vector part of the moduli space depending on scalar fields $z, \bar z$. It is, however, important to realize that we will only use here (in effective N=1 supergravity) the fact that the explicit expression for ${\cal N}$ is defined by the section and the derivative of it over the moduli.

One can introduce a symplectic  charge related to the integer flux in a compactified Calabi-Yau manifold:
$(
p^\Lambda  ,  q_\Lambda)
$.
Now  we may define a {\it covariantly holomorphic central change} 
$$
Z(z, \bar z, q,p) \equiv  (L^\Lambda
q_\Lambda -
M_\Lambda p^\Lambda) \ ,
\label{central}$$ 
where $D_{\bar i} Z\equiv (\partial_{\bar i} - {1\over 2} K_{
\bar i}) Z =0$ and $D_i \bar Z\equiv (\partial_{ i} - {1\over 2} K_{
 i}) \bar Z = 0$.
One could also define a {\it holomorphic central charge}  
$$
W= e^{-K(z, \bar z)/2} Z(z, \bar z, q,p) \equiv  (X^\Lambda
q_\Lambda -
F_\Lambda p^\Lambda) \ , \qquad \partial_{ \bar i}W=0\ .
\label{superpotential}$$ 
This holomorphic charge $W$ may be associated with  the superpotential in N=1 effective supergravity for the IIB string theory compactified on CY orientifold.
In the generic point of the moduli space there are  two symplectic    invariants homogeneous of degree 2 in electric and magnetic charges:
\begin{equation}
I_1 =  I_1(p,q,z,\bar z)=-{1\over 2} P^t {\cal M}({\cal N}) P\ ,
\qquad
I_2 = I_2(p,q,z,\bar z)=-{1\over 2} P^t {\cal M}({\cal F}) P \ . 
\label{F}
\end{equation}
Here $P=(p,q)$ and ${\cal M}({\cal N})$ is the real symplectic $(2n+2) \times (2n+2)$ matrix
$$
\pmatrix{
{\rm Im} {\cal N} + {\rm Re} {\cal N} {\rm Im} {\cal N}^{-1} {\rm Re} {\cal N}\ &&& - {\rm Re} {\cal N} \,
{\rm Im} {\cal N}^{-1} \cr
-{\rm Im} {\cal N}^{-1}  {\rm Re} {\cal N}  \ &&&{\rm Im} {\cal N}^{-1}  \cr
}
$$
The matrix ${\cal M}({\cal F})$ is an analogous matrix with ${\cal N}$ replaced by ${\cal F}$, where ${\cal F}= \partial_\Lambda F_\Sigma $.
Using the central charge one can rewrite these two invariants as follows:
\begin{equation}
I_1 =  |Z|^2 + |D_i Z|^2 \ , \qquad
I_2 =  |Z|^2 - |D_i Z|^2 \ .
\label{inv}
\end{equation}
An effective N=1 d=4 supergravity potential from IIB string theory on CY orientifold given in eq. (\ref{pot}) in a generic point of the moduli space can be also presented in the form
\begin{equation}
  V= -2I_2-I_1 \ ,
\label{symplectic}
\end{equation}
where $I_1$ and $I_2$ are defined in eqs. (\ref{inv}).
An extremization condition for both symplectic invariants, which specifies the  values of moduli in terms of charges,  is given by 
$$
D_i Z(z, \bar z, q,p) \equiv (\partial_{ i} + {1\over 2} K_{
i}) Z =0 \ .$$ 
It is also a requirement of an unbroken supersymmetry.
In terms of the holomorphic charge $W$, the extremization condition is even more familiar:
$$
 D_i W(z,q,p)\equiv (\partial_{ i} +  K_{
i}) W =0 \ . $$ 
Note that $X^\Lambda(z)$ are subject to holomorphic redefinitions (sections of a holomorphic line bundle):
$$\label{xx1}
X^\Lambda(z) \to X^\Lambda(z)~e^{-f(z)} \ ,
$$
so that $$\label{xx2}
L^\Lambda(z) \to  L^\Lambda(z)~e^{\bar f(z) -f(z)\over 2}\  .
$$
This occurs because $L^\Lambda = e^{K/2} X^\Lambda$ and $K\rightarrow K+f+\bar f$
under  K\"ahler transformations, so that 
\begin{equation}
 Z(q,p,z, \bar z) \to Z(q,p,z, \bar z)~e^{\bar f(\bar z) -f(z)\over 2} \ .
\label{xx3}
\end{equation}
However,  $|Z|$ is both symplectic and K\"ahler gauge invariant, this is why the connection drops  and $D_i Z = 0$ ($D_{\bar i}Z \equiv 0$) means that $\partial_i |Z| = 0$.  At the attractor point, $D_i Z=0$, the potential  does not depend on moduli, only on fluxes:
$$
 V_{cr}=-3 I_1= -3I_2= -3 |Z(\bar z(p,q), z(p,q),q,p)|^2 \ .
\label{attractorZ}
$$
 The critical point of the potential at $D_i Z=0$ can be also presented in the form of the  attractor equations
\begin{equation}
 \left (\matrix{
p^\Lambda\cr
q_\Lambda\cr
}\right )= i \left(\matrix{
\bar Z L^\Lambda- Z\bar \Lambda^\Lambda\cr
 \bar Z M_\Lambda- Z \bar M_\Lambda\cr
}\right ) \ .
\label{stab}
\end{equation}
From  these equations it follows that $(p,q)$  determine the sections up to a
(K\"ahler) gauge transformation. The moduli at the fixed point depend on ratios of charges since the equations  are homogeneous in $p,\,q$.

For completeness we present here the derivation of these equations following \cite{Ferrara:1996dd}. It will be clear that the special geometry of the moduli space present in N=1 supergravity derived from string theory on CY orientifold is sufficient to derive these equations. Start with
\begin{equation}
D_{\bar i}\bar Z   =  D_{\bar i}\bar L^\Lambda q_\Lambda -  D_{\bar i}\bar M_\Lambda p^\Lambda = 0 \ .
\label{dif}
\end{equation}
Now replace $D_{\bar i}\bar M_\Lambda$ by ${\cal N}_{\Lambda\Sigma}D_{\bar i}\bar L^{\Sigma}$, according to eq. (\ref{Ncal}).
By contracting with $D_iL^\Sigma G^{i\bar i}$ and using the identity
$
D_iL^\Sigma G^{i\bar i} D_{\bar i}\bar L^\Lambda = - {\textstyle {1\over 2}} {{\rm Im}} {({\cal N}^{-1})}^{\Sigma\Lambda} - \bar L^{\Sigma}L^\Lambda 
$
we get 
\begin{equation}\label{u2}
2Z \bar L^\Sigma = i p ^\Sigma - {{\rm Im}} {({\cal N}^{-1})}^{\Sigma\Lambda}\, q_\Lambda + {{\rm Im}}  {({\cal N}^{-1})}^{\Sigma\Gamma}\  {{\rm Re}}   {\cal N}_{\Gamma\Delta}\  p^\Delta  \ ,
\end{equation}
from which it follows that
\begin{equation}\label{u3}
  p^{\Sigma}  =  2i\bar Z  L^\Sigma   - i  ({{\rm Im}}  {\cal N}^{-1}\  {{\rm Re}}  {\cal N}\, p  +  {\rm Im} {\cal N}^{-1} \, q )^\Sigma \ ,
\end{equation}
\begin{equation}\label{u4}
q_{\Sigma} = 2i\bar Z  M_\Sigma     -i  ({{\rm Im}}  {\cal N} \, p  + {{\rm Re}}  {\cal N}\    {{\rm Im}} {{\cal N}^{-1} }\  {{\rm Re}}\,  {\cal N} \, p - {{\rm Re}} {\cal N}\  {{\rm Im}}  { {\cal N}^{-1} }\, q )_\Sigma\ .
\end{equation}
Now we see that it is really important to have a symplectic charge $(p^\Lambda, q_\Lambda)$ which is real and moduli independent: note that we have differentiated  in eq. (\ref{dif}) over the moduli inside $\bar Z$ only the section  $\bar L^\Lambda, \bar M_\Lambda$,  not the flux $p,q$. In such case we find from eqs. (\ref{u3}) and (\ref{u4}) the attractor eqs. (\ref{stab}).

This detailed derivation of the attractor equations  explains why the use of the $G_3=F_3-\tau H_3$ flux in type IIB theory  obscures the derivation of the attractor equations for the flux vacua. It also suggests that one should try to rewrite the potential for N=1 effective supergravity in IIB theory compactified on a CY orientifold in a form suitable for the application of attractor equations: {\it one should  use only the quantized fluxes not mixed with moduli, and one should push all dependence on all moduli into the properly defined symplectic section.}

\section{ Type IIB Flux Vacua as Attractors} 

Consider  a compactification of IIB string theory on some CY 3-fold with O3/O7 planes. We refer the reader to a nice recent review of the related topics in \cite{Grana:2005jc}. The 3-cycles come in pairs (A, B) so that (with $(2\pi)^2 \alpha'=1$) the RR 3-form $F_3$ and the NS 3-form $H_3$ are 
\begin{equation}
  F_3= p_f^a \,\alpha_a - q_{af} \,\beta^a \ , \qquad  H_3= p_h^a \,\alpha_a - q_{ah}\, \beta^a \ ,
\label{H3}
\end{equation}
 where 
\begin{equation}
  \int_{CY} \alpha_a\wedge \beta^b= \delta_a{}^b \ .
\label{norm}
\end{equation}
The holomorphic symplectic basis on CY is such that
$$
  X^a(x)= \int_{A^a} \Omega \qquad G_a(x)= \int_{B_a} \Omega \ ,
\label{holomorphic} 
$$
and 
\begin{equation}
  \Omega_3= X^a(x)\, \alpha_a - G_{a}(x)\, \beta^a \ .
\label{H3b}
\end{equation}
Here $ \Omega(x)$ is a  holomorphic 3-form on a Calabi-Yau space depending on the complex structure moduli $x$, and the K\"ahler potential of CY is given by
\begin{equation}
 K(x, \bar x)= - \ln [ i\int  \Omega(x)\wedge \bar \Omega(\bar x)]= - \ln i[\bar X^a G_a-X^a\bar G_a]  \ .
\label{K}
\end{equation}
This standard description of the compactified IIB string theory is supplemented by the axion-dilaton moduli as follows. From the $SL(2,Z)$ doublet of fluxes, $F$ and $H$, one forms a complex 3-form $G_3= F_3-\tau H_3$, and the K\"ahler potential has an additional part, so that the total K is
\begin{equation}
  K(\tau, \bar \tau, x, \bar x)=- \ln [-i (\tau-\bar \tau)]- \ln [ i\int  \Omega(x)\wedge \bar \Omega(\bar x)] \ .
\label{dialtonkahler}
\end{equation}
The superpotential is given by
\begin{equation}
  W= \int G_3(\tau)\wedge \Omega(x)= (p_f^a -\tau p_h^a )G_a - (q_{fa} -\tau q_{ha})X^a \ .
\label{W}
\end{equation}
It is this particular dependence on the axion-dilaton that makes the symplectic structure of the flux vacua obscure.

Let us instead make the $SL(2, Z)$ symmetry of the type IIB theory manifest, so that it extends the manifest symplectic symmetry of the CY space. The first hint comes from rewriting the superpotential as follows: 
\begin{equation}
  W= \int G_3(\tau)\wedge \Omega(x)= p_f^a G_a- q_{fa}X^a+
  p_h^a (-\tau  G_a)   -  q_{ha}(- \tau X^a) \ .
\label{W1}
\end{equation}
Let us now introduce  the ``central charge'' in a form useful for the attractor equation: 
\begin{equation}
Z=e^{K(x, \bar x, \tau, \bar \tau)\over 2}  W (\tau, x)= e^{K(x, \bar x, \tau, \bar \tau)\over 2}\int F_3\wedge \Omega + H_3\wedge (-\tau \Omega) \ .
\label{Z}
\end{equation}
Thus we have a 3-form flux $SL(2,Z)$ doublet
\begin{equation}
 F= (F_3, \, H_3) \ ,
\label{flux doublet}
\end{equation}
and a symplectic section, which is also a doublet,
\begin{equation}
\Pi= \left (\matrix{
\Pi_1(\tau, x)\cr
\Pi_2(\tau, x)\cr
}\right )
  = \left (\matrix{
\Omega(x)\cr
-\tau \Omega(x)\cr
}\right ) \ .
\label{sectiondoublet}
\end{equation}
The total K\"ahler potential is now given by
\begin{eqnarray}
K(x, \bar x, \tau, \bar \tau )&=& - \ln \left[ \int [ \tau \Omega(x)\wedge \bar \Omega(\bar x)-  \Omega(x)\wedge \bar \tau \bar \Omega(\bar x)]\right]\nonumber\\
&=& -\ln \left[\int (\Pi_1\wedge \bar  \Pi_2 -  \Pi_2\wedge \bar \Pi_1)\right] \ .
\label{Kahler}
\end{eqnarray}
The central charge is
\begin{equation}
Z=e^{K\over 2}  W = e^{K\over 2} \int F\wedge \Pi \ .
\label{newZ}
\end{equation}

\subsection{$SL(2,Z)$ symmetry}

Under $SL(2,Z)$ transformations with
\begin{equation}
   \tau'= {a\tau +b\over c\tau +d}
\label{R}
\end{equation}
the flux doublet transforms as 
\begin{equation}
  \left (\matrix{
F_3\cr
H_3\cr
}\right )'= R   \left (\matrix{
F_3\cr
H_3\cr
}\right )  \ , \qquad R= \left (\matrix{
a & b\cr
c& d\cr
}\right )\ .
\label{fluxtransf}
\end{equation}
The central charge, defined in eqs. (\ref{Z}), (\ref{newZ}) transforms with the phase
\begin{equation}
  Z' = e^{-i Arg(c\bar \tau+d)} Z \ ,
\label{Ztransf}
\end{equation}
as was shown in studies of axion-dilaton black holes in \cite{Kallosh:1993yg}.
This is clearly compensated by the K\"ahler transformation of the type (\ref{xx3}).
This concludes the derivation of the  special geometry in the moduli space $z=(\tau, x)$ of the axion-dilaton $\tau$ and the complex structure fields $x$ of the CY$_3$. The attractor equations for the flux vacua of type IIB string theory on CY$_3$ orientifold can be guessed by analogy with black hole attractor equations, at least for the case that the special geometry rule is satisfied, i. e. at the fixed point $DZ=0$ with additional requirement that $M_{AB}= D_A D_B Z=0$
\begin{equation}
  \left (\matrix{
p_h^a\cr
\cr
q_{ha}\cr
\cr
p_f^a\cr
\cr
q_{fa}\cr
}\right )=   e^{K} \left (\matrix{
 \, \bar W   X^a + \; W \bar X^a\cr
\cr
 \, \bar W   G_a+\; W \bar G_a \cr
\cr
 \tau \bar W   X^a+\bar \tau   W X^a \cr
\cr
 \tau \bar W   G_a+\bar \tau  W \bar G_a\cr
}\right )_{fix}
\label{IIBmassless}
\end{equation}
Here the right hand side has explicit dependence on the universal axion-dilaton and a generic dependence on the complex structure moduli of an arbitrary CY$_3$. All moduli in the rhs. of this equation take fixed values defined by fluxes in the lhs. of the attractor equation.  


\subsection{Simplified notation}

We may also introduce the form notation to simplify the axion-dilaton dependence.
Let  
\begin{equation}
  F_4= - \alpha\wedge F_3 +\beta \wedge H_3 \ , \qquad  \int_{T^2}\alpha\wedge \beta =1 \ .
\label{torus}
\end{equation}
The complex structure of the auxiliary torus is $\omega= \beta - \tau \alpha$. We can define the holomorphic 4-form
\begin{equation}
\Omega\equiv \Omega_{(0,4)}(z)= \Omega_{(0,3)}(x)\wedge\omega(\tau)= \Omega_{(0,3)}\wedge \beta - \tau \Omega_{(0,3)} \wedge \alpha \ .
\label{4period}
\end{equation}
The total K\"ahler potential is now
\begin{equation}
 K (t, \bar t)=  -\ln  \int_{X_3\times T^2}\Omega_{(0,4)}\wedge \bar \Omega_{(4,0)} \ .
\label{Ktorus}
\end{equation}
The covariantly holomorphic central charge and the superpotential are
\begin{equation}
  Z= e^{K\over 2} \int_{X_3\times T^2} F_4\wedge \Omega_{(0,4)} \ ,  \qquad W= \int_{X_3\times T^2} F_4 \wedge \Omega_{(0,4)} \ .
\label{central2}
\end{equation}
The flux attractor equations take a  remarkably simple form
\begin{equation}
F_4= [e^{K} (\bar W \Omega + W \bar \Omega)]_{fix}= 2 e^{K} \rm Re \, ( W \, \bar \Omega_{(4,0)})|_{fix} \ .
\label{attractorsimple}
\end{equation}
We may also present it using the central charge and covariantly holomorphic form $\hat \Omega_4 = e^{K\over 2}\Omega_{(0,4)}$
\begin{equation}
F_4=  [\bar Z \hat \Omega + Z \hat {\bar \Omega}]_{fix}= 2  \rm Re \, ( Z \, \hat {\bar  \Omega})|_{fix} \ .
\label{attractorsimple1}
\end{equation}
This is a condensed form of equation (\ref{IIBmassless}).
The flux vacua attractor equation (\ref{attractorsimple1}) is almost the same equation as the black hole attractor equation. The difference is that in black hole case we have $\rm Im (Z\hat {\bar \Omega}_3)$ defining the 3-form flux whereas in flux vacua case we have $\rm Re (Z\hat {\bar \Omega}_4)$ defining the effective 4-form flux.

\subsection{Example of M-theory on K3$\times$K3}
As an example, we will consider here 
stabilization of the complex structure moduli in M-theory on K3$\times$K3 \cite{Aspinwall:2005ad,Moore:2004fg}, which is also related to stabilization of the complex structure in IIB string theory on K3$\times T^2\over Z_2$. For the purpose of relating flux vacua with black hole attractors, we will slightly modify the procedure used in \cite{Aspinwall:2005ad}, where we were interested in stabilizing all moduli in this model. For this purpose we  made a choice of the 4-form flux as a $(2,2)$ form. This is a choice when the superpotential defined by eq. (\ref{central2}) vanishes. Here we will fix the complex structure moduli by fluxes in such a way that the superpotential does not vanish: we will choose the real 4-form flux as follows
\begin{equation}
  F_4= \bar c \,
  \Omega^1\wedge \Omega^2+ c \,\bar \Omega^1\wedge \bar \Omega^2 \ .
\label{4flux}
\end{equation}
Here $\Omega^i$,  $i=1,2$ are holomorphic 2-forms on each of K3 at the attractor point.
The superpotential can be calculated, and we find 
\begin{equation}
   W= \int_{K3\times K3}  c \bar \Omega^1\wedge \bar \Omega^2 \wedge \Omega^1\wedge\Omega^2 =  c \, e^{-K}\ , \qquad \Rightarrow \qquad c=Z
\label{super}
\end{equation}
This proves that we may now rewrite the 4-form flux as follows
\begin{equation}
  F_4= 2 e^K \rm Re (\bar W \,
  \Omega^1\wedge \Omega^2) = 2\rm Re (\bar Z \,
  \hat \Omega^1\wedge \hat \Omega^2)  \ .
\label{4fluxattr}
\end{equation}
Since in this example $\Omega_{(0.4)}= \Omega^1\wedge \Omega^2$, we  confirm for this example the   flux vacua attractor formula (\ref{attractorsimple}), (\ref{attractorsimple1}) derived in the general case for type IIB string theory compactified on a generic CY orientifold.

We may now take $\Omega_i, \, i=1,2$ proportional to $ p_i+\tau_i q_i$, as suggested in \cite{Aspinwall:2005ad,Moore:2004fg}. 
Two complex numbers $\tau_j$ for this choice of the holomorphic form  are fixed by the condition
$\Omega_j^2=0$ to be
$
  \tau_j = (-p_j.q_j + i\sqrt{\det Q_j})/{q_j^2}$. Each of the attractive K3 surfaces is now described up to SL(2,Z) equivalence class by a matrix \cite{Aspinwall:2005ad}
\begin{equation}
 Q_j= \left (\matrix{
p_j^2&p_j.q_j\cr
\cr
p_j.q_j&q_j^2\cr
}\right )
\end{equation}\label{qi2}
 (no summation over $j$).                     
In  the definition of the attractive K3 surface  in \cite{Moore:2004fg} it was pointed out that  ${A\over 4\pi}=|Z|_{fix}^2= \sqrt{ Det \, Q}$
is an area of the unit cell in the transcendental lattice $T_S$ of the K3 surface. 
As we will see now, our example  helps to reveal another interesting relation between black holes and flux vacua: a relation between the black hole entropy and the transcendental lattice  $T_S$ on the attractive K3.

Indeed, 
the area of the horizon of the supersymmetric black holes in ${SU(1,1)\over U(1)} \times {SO(2,n)\over SO(2)\times SO(n)}$ symmetric
manifold was established in \cite{Kallosh:1996tf}:
\begin{equation}
  {A\over 4\pi}= |Z|_{fix}^2= \sqrt{ Det \, Q}=  (p^2 q^2- (p\cdot q)^2)^{1/2} \ .
\label{blackhole}
\end{equation}
This result was found by solving explicitly the supersymmetric black hole attractor equations in this theory and stabilizing all moduli of the ${SU(1,1)\over U(1)} \times {SO(2,n)\over SO(2)\times SO(n)}$ coset space in terms of the black hole charges. Now   one finds   the same  value of $|Z|_{fix}^2$  in the context of flux vacua for each K3.  We leave the detailed study of this issue for the future publication.

Note that Eq. (\ref{attractorsimple}), which in detailed form is given in Eq. (\ref{IIBmassless}), was derived either using the tools of special geometry or using the tools of algebraic geometry such as Hodge-decomposition  of the space of allowed 4-form fluxes.  It is valid for generic CY$_3$. As we have shown in particular example of K3$\times {T^2/Z_2}$ and its M-theory version, the coefficient of proportionality between fluxes and period matrix is linear in the critical value of the superpotential (central charge).

\section{Conclusion}

In conclusion, we have established that a class of flux vacua with $DW=0$ and $W\neq 0$ are defined by the attractor equations (\ref{attractorsimple}), (\ref{IIBmassless}), which have the same form as the attractor equations for the supersymmetric black holes near the horizon. Note that these vacua with $W_0(z)_{fix}\neq 0$ have been used in the KKLT construction \cite{Kachru:2003aw}.  We have
used the symplectic structure of the type IIB string theory compactified on CY orientifold. The corresponding symmetry of the moduli space is a product of the $Sp(2n, Z)$ symmetry and $SL(2,Z)$ symmetry. The first one originates from the special geometry of  the CY space, whereas the second one comes from the $SL(2,Z)$ symmetry of the type IIB theory in d=10. It is most easily expressed via the special geometry of the product space of the CY$_3$-fold and an auxiliary torus $T^2$.  Using the covariantly holomorphic central charge, $Z=e^{K/2}$, we have shown that for the non-vanishing value of the central charge equation $D_iZ=0$ can be replaced by an attractor equation $F=2\rm Re(\bar Z 
\hat \Omega)$. We should stress, however, that there is in general a deviation from the special geometry which is present in flux vacua, in general, as $D_i D_j Z$ may not vanish: only when it is vanishing the vacua are defined by the black hole attractor equation, see   
  \cite{Kallosh:2005ax}.

In case of regular supersymmetric black holes in classical N=2 supergravity,  the square of the central charge  $|Z|^2$ at the attractor point is proportional to the black hole entropy/area of the horizon, which at the attractor point depends only on charges and does not depend on continuous moduli. Typically the entropy is given by a product of charges. If one of the charges in this product vanishes, the area of the horizon vanishes, the metric has a null singularity, and some of the moduli may blow up. It was therefore important in the studies of supersymmetric black holes near the horizon to find cases with  central charge which does not vanish at the attractor point. The function which was minimized was $|Z(z, \bar z, p, q)|^2$ with $z, \bar z$ taking arbitrary values in the moduli space corresponding to the values of these moduli at the asymptotic infinity. At the minimum, this function was proportional to the area of the horizon, $ {A(p, q)\over 4\pi}=  |Z(z(p,q), \bar z(p,q), p, q)|^2$.

In flux vacua we are looking for the critical points of the potential $V= |DZ|^2- 3 |Z|^2$. The cases of flux vacua with $DZ=0$,  $Z=0$ and $V_{fix}=0$ have solutions with all moduli stabilized  (axion-dilaton and complex structure), as different from the situation with extremal black holes\footnote{Extremal black holes with $Z=0$ require some special treatment, see e. g. \cite{Denef:2001xn}, \cite{Dabholkar:2004dq}.}. 
We will show in  \cite{Kallosh:2005ax} that such vacua can be described by the generalized  attractor equations. Here we have shown that at the critical point $DZ=0$, $Z\neq  0$ the potential  is equal to $V=-3m_{3/2}^2=-3|Z|^2$. The relation between charges and fixed moduli in this case can be described by the black hole attractor type equation.  This is the main point of this note: flux vacua defined by differential equations $D_i Z= D_{i} D_{j} Z=0$ and the condition $Z\neq 0$ can be described by the  algebraic equations analogous to  supersymmetric black holes attractors.

 \

{\it Note added}. The main content of this paper is now included into a new extended version of ref. \cite{Kallosh:2005ax},  following the suggestion of the referee of JHEP.

\ 

\leftline{\bf Acknowledgments}

I am  grateful to P. Aspinwall, F. Denef, S. Ferrara, 
S. Kachru, A. Linde, M. Sasaki,  S. Trivedi, P. Yi and D. Waldram for
valuable discussions of flux vacua and attractors. I am grateful for the hospitality to the organizers of the conference ``The Next Chapter in Einstein's Legacy'' at the Yukawa Institute, Kyoto, where this work was performed.
It was supported by NSF grant 0244728 and by Kyoto University.

\end{document}